\documentstyle[11pt,mrs2001,epsfig]{article}

\begin{document}
\newcommand{\hMsun}{{h^{-1}}{\rm M}_{\sun}}
\newcommand{\hmsun}{{h^{-1}}{\rm M}_{\sun}}
\newcommand{\hmpc}{\ifmmode{h^{-1}{\rm Mpc}}\;\else${h^{-1}}${\rm Mpc}\fi}
\newcommand{\hkpc}{\ifmmode{h^{-1}{\rm kpc}}\;\else${h^{-1}}${\rm kpc}\fi}
\newcommand{\lcdm}{$\Lambda$CDM}
\newcommand{\msun}{{\rm M}_{\solar}}
\newcommand{\LCDM}{$\Lambda$CDM}
\newcommand{\rhonfw}{\rho_{\sss \rm NFW}}
\newcommand{\rhos}{\rho_{\rm s}}
\newcommand{\sss}{\scriptscriptstyle}
\newcommand{\cvir}{\ifmmode{c_{\rm vir}}\else$c_{\rm vir}$\fi}
\newcommand{\rs}{\ifmmode{R_{\rm s}}\else$R_{\rm s}$\fi}
\newcommand{\Rvir}{\ifmmode{R_{\rm vir}}\else$R_{\rm vir}$\fi}
\newcommand{\ac}{\ifmmode{a_{\rm c}}\else$a_{\rm c}$\fi}
\newcommand{\cc}{\ifmmode{c_{\rm 1}}\else$c_{\rm 1}$\fi}
\newcommand{\zf}{\ifmmode{z_{\rm f}}\else$z_{\rm f}$\fi}
\newcommand{\zc}{\ifmmode{z_{\rm c}}\else$z_{\rm c}$\fi}
\newcommand{\zobs}{\ifmmode{z_{\rm o}}\else$z_{\rm o}$\fi}
\newcommand{\aobs}{\ifmmode{a_{\rm o}}\else$a_{\rm o}$\fi}
\newcommand{\Mobs}{\ifmmode{M_{\rm o}}\else$M_{\rm o}$\fi}
\newcommand{\beq}{\begin{equation}}
\newcommand{\eeq}{\end{equation}}

\title{Concentrations and Assembly Histories of Dark Matter Halos}
\author{R. H. WECHSLER$^{1,2}$, J. S. BULLOCK$^{3}$, J. R. PRIMACK$^{1}$, A. V. KRAVTSOV$^{3,4}$, \& A. DEKEL$^5$}
\affil{$^1$Physics Department, University of California, Santa Cruz, CA, 95064}
\affil{$^2$Physics Department, University of Michigan, Ann Arbor, MI 48109}  
\affil{$^3$Department of Astronomy, The Ohio State University, Columbus, OH 43210}
\affil{$^4$Hubble Fellow}
\affil{$^5$Racah Institute of Physics, The Hebrew University, Jerusalem 91904 Israel}

\begin{abstract}
We study the relation between the density profiles of dark matter
halos and their mass assembly histories, using a statistical sample of
halos in a high-resolution N-body simulation of the \LCDM\ cosmology.
For each halo at $z=0$, we identify its merger-history tree, and
determine concentration parameters $\cvir$ for all progenitors, thus
providing a structural merger tree for each halo.  We fit the mass
accretion histories by a universal function with one parameter, the
formation epoch \ac, defined when the log mass accretion rate ${\rm
d}\log M/{\rm d}\log a$ falls below a critical value.  We find that
late forming galaxies tend to be less concentrated, such that $\cvir$
``observed'' at any epoch $\aobs$ is strongly correlated with $\ac$
via $\cvir=\cc \aobs /\ac$.  Scatter about this relation is mostly due
to measurement errors in $\cvir$ and $\ac$, implying that the actual
spread in $\cvir$ for halos of a given mass can be mostly attributed
to scatter in $\ac$.  Because of the direct connection between halo
concentration and velocity rotation curves, and because of probable
connections between halo mass assembly history and star formation
history, the tight correlation between these properties provides an
essential new ingredient for galaxy formation modeling.

\end{abstract}

\section{METHOD}

We investigate the connection between halo density profiles and their
mass assembly histories, using a structural merger tree constructed
from a high-resolution N-body simulation of a flat \LCDM\ model with
$\Omega_m = 0.3$, $h = 0.7$ and $\sigma_8$ = 1.0, whose evolution has
been simulated with the ART code \cite{kkk:97}. The trajectories of
$256^3$ cold dark matter particles are followed within a cubic,
periodic box of comoving size 60\hmpc\ from redshift $z = 40$ to the
present.  We use distinct halo catalogs at 36 output times spaced
between $z=7$ and $0$.  NFW density profiles \cite{nfw:96},
$\rhonfw(r) = \rhos/[(r/\rs)\left(1+r/\rs\right)^2]$, are measured for
each halo with more than 200 particles, corresponding to halos more
massive than $2.2\times10^{11}\hMsun$.  For each halo at $z=0$, we
identify its full merger history, and determine concentration
parameters $\cvir\equiv\Rvir/\rs$ for all progenitors, thus providing
a structural merger tree for each of $\sim 3000$ halos.

\begin{figure}
\plottwo{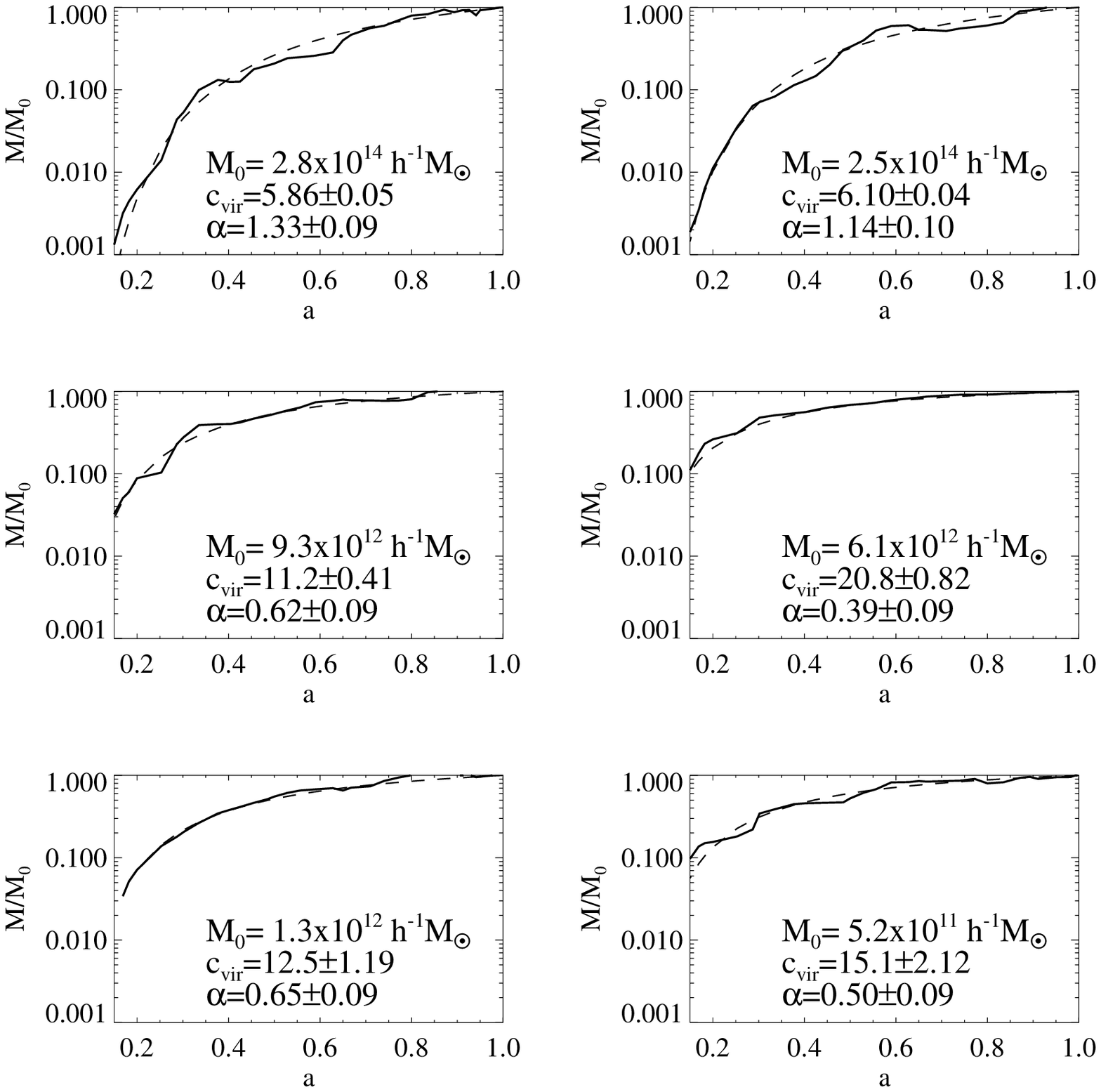}{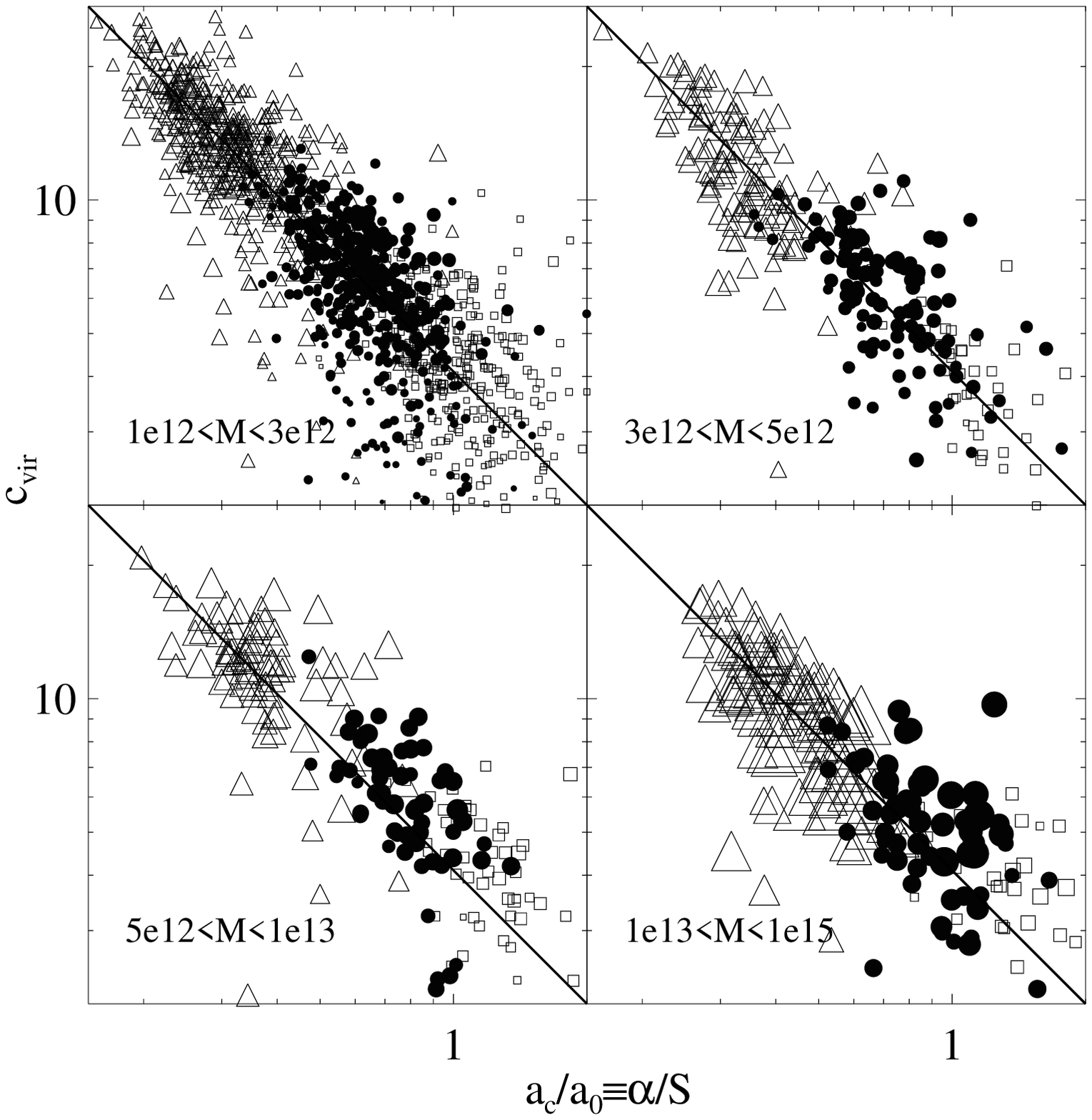}
\caption	
{{\bf Left:} Selected mass accretion trajectories, showing the
evolution of the most massive progenitor for individual halos in the
simulation (thick).  Functional fits to the growth curve of each halo
using Eq. \ref{eq:fit} are shown as thin smooth lines.  {\bf Right:}
Concentration versus scaled formation epoch $\ac/\aobs$, for halos at
$z=0$ (triangles), $z=1$ (circles), and $z=2$ (squares).  The 4 panels
correspond to different mass ranges.  At all masses and redshifts, the
concentration parameter $\cvir$ is well fit by the functional form $\cvir
= \cc\ac/\aobs$, where $\cc \sim 4.1$ (represented by the solid line
in each panel).
\label{figure}
}
\end{figure}

\section{RESULTS}
Figure~\ref{figure} (left) shows the history of mass growth for the
major progenitors of several different halos, spanning a range of
masses and concentration parameters.  Massive halos tend show
substantial mass accumulation up to late times, while the growth
curves for less massive halos tend to flatten out earlier.  By
examining a range of full mass assembly histories for our sample of
halos, we find that both average mass accretion histories and mass
accretion histories for individual halos are well characterized by a
simple function: $M(a) = \Mobs e^{-\alpha z}, \quad a=(1+z)^{-1}$.
Fits to this equation are shown in Figure
\ref{figure} (left) for representative individual halos.
The single free parameter $\alpha$ can be related to a characteristic
epoch for formation, \ac, defined as the expansion scale factor $a$
when the logarithmic slope of the accretion rate, ${\rm d}\log M/ {\rm
d}\log a$, falls below some specified value, $S$; the functional form
implies $\ac = \alpha/S$.  The same formation epoch can be defined
equivalently for any ``observing'' epoch $\zobs$ of that halo, by
replacing $a$ by $a/\aobs$, in which case $\ac =\aobs\alpha/S$.  Thus
at any such observing redshift, with scalefactor $\aobs=1/(1+\zobs)$
and mass $\Mobs=M(\zobs)$, the mass growth is fit by
\beq
M(a) = \Mobs {\rm exp} \left[-\ac S \left(\frac{\aobs}{a}-1\right)\right].
\label{eq:fit}
\eeq
This implies that, for any halo whose mass accretion trajectory follows
this functional form, the characteristic formation time is the
same regardless of the redshift $\zobs$ at which the halo is observed.

We find that the concentration of a halo, defined as $\cvir \equiv
\Rvir/\rs$, is tightly correlated with the characteristic formation
epoch as defined above, and that this relation holds at all redshifts
when, properly scaled by $\aobs$:
\beq
\cvir = \cc \aobs/\ac,  
\label{eq:c}
\eeq
where $\cc$ is the typical concentration of halos whose formation time
is at the time of measurement, $a_c=\aobs$.  Figure
\ref{figure} (right) shows that this formula provides a good description
of the observed correlation between concentration and formation time
for halos at all masses and redshifts.  As mentioned, the typical
formation time is a function of mass, but there is significant scatter
in $\ac$ for a fixed mass.  The relation defined by Eq. \ref{eq:c} is
able to account for the complete mass and redshift dependence of
$\cvir$, and for the scatter in $\cvir$ measured for fixed mass halos.
This correlation is has important consequences for galaxy formation
modeling.  For further details on this work, see \cite{rw:01}.

\vfill
\end{document}